\documentclass[a4paper]{ifacconf}

\usepackage[round]{natbib}             
\usepackage{graphicx,amsmath,url} 
\usepackage{caption}
\usepackage{subcaption}
 \def\portugues{0} 
\ifx\portugues\undefined
\def\portugues{0}
\fi

\if\portugues0
   \usepackage[english]{babel}
  \else
   \usepackage[spanish,brazil,english]{babel}
\fi

\usepackage[T1]{fontenc}

\usepackage[utf8]{inputenc}

\usepackage{ae}

\if\portugues1
 { }
 { }
 { }
 
\fi

\begin{document}

\if\portugues1

%
\selectlanguage{brazil}
	
\begin{frontmatter}

\title{Estilo para Artigos em Português das Conferências da SBA --- Adaptado do  IFAC\thanksref{footnoteinfo}} 

\thanks[footnoteinfo]{Reconhecimento do suporte financeiro deve vir nesta nota de rodapé.}

\author[First]{Primeiro A. Autor} 
\author[Second]{Segundo B. Autor} 
\author[Third]{Terceiro C. Autor}

\address[First]{Faculdade de Engenharia Elétrica, Universidade do Triângulo, MG, (e-mail: autor1@faceg@univt.br).}
\address[Second]{Faculdade de Engenharia de Controle \& Automação, Universidade do Futuro, RJ (e-mail: autor2@feca.unifutu.rj)}
\address[Third]{Electrical Engineering Department, 
   Seoul National University, Seoul, Korea, (e-mail: author3@snu.ac.kr)}

\selectlanguage{english}
\renewcommand{\abstractname}{{\bf Abstract:~}}
\begin{abstract}                
These instructions give you guidelines for preparing papers for the Sociedade Brasileira de Automática (SBA) technical meetings using the IFAC style. Please use this document as a template to prepare your manuscript in portuguese. For submission guidelines, follow instructions on paper submission system as well as the event website.

\vskip 1mm
\selectlanguage{brazil}
{\noindent \bf Resumo}:  As instruções abaixo são linhas gerais para a preparação de artigos para conferências e simpósios da Sociedade Brasileira de Automática (SBA) usando como base o estilo IFAC. Instruções de submissão podem ser encontradas no sistema de submissão de artigos ou no {\em website} do congresso.
\end{abstract}

\selectlanguage{english}

\begin{keyword}
Five to ten keywords separatety by semicolon. 

\vskip 1mm
\selectlanguage{brazil}
{\noindent\it Palavras-chaves:} Utilize de cinco a dez palavras-chaves separadas por ponto e vírgula.
\end{keyword}

\selectlanguage{brazil}

\end{frontmatter}
\else
%

\begin{frontmatter}

\title{RTI-NMPC for Control of Autonomous Vehicles Using Implicit Discretization Methods} 

\author[First]{Matheus Wagner} 
\author[Second]{Julio E. Normey-Rico} 

\address[First]{Dept. de Automação e Sistemas ( DAS), Univ. Fed. de Santa
Catarina, Florianópolis-SC, Brazil. (email: matheus.wagner@posgrad.ufsc.br)}
\address[Second]{Dept. de Automação e Sistemas ( DAS), Univ. Fed. de Santa
Catarina, Florianópolis-SC, Brazil. (email: julio.normey@ufsc.br)}

\renewcommand{\abstractname}{{\bf Abstract:~}}   
   
\begin{abstract}  
Recent efforts in the development of autonomous driving technology have induced great advancements in perception, planning and control systems. Model predictive control is one of the most popular advanced control methods, but its application to nonlinear systems still depends on the development of computationally efficient methods. This work presents a nonlinear model predictive control formulation based on real-time iteration using an implicit discretization of the system's dynamics, with the objective of achieving greater prediction accuracy and lower computational cost when dealing with stiff dynamical systems, as is the case for vehicle dynamics. The proposed method is described and later evaluated on a simulation scenario considering modeling errors and external disturbances. The presented  results demonstrate the effectiveness of the method when it comes to tracking a given trajectory and its low computational burden, measured in terms of execution time.
\end{abstract}

\begin{keyword}
Nonlinear predictive control, Motion Control Systems, Guidance navigation and control, Autonomous Vehicles, Trajectory Tracking and Path Following
\end{keyword}

\end{frontmatter}
\fi


\section{Introduction}
In recent times major research efforts, from both academia and the private sector, have been put into the development of autonomous driving technology, as it has the potential to improve the overall safety, efficiency and accessibility of transportation systems \citep{yu2021model}.

Autonomous driving systems require the interaction between several technologies. These technologies include perception, motion planning and control systems, the latter being responsible for realizing the planned trajectory by estimating the vehicle's state and using it to define which actuation commands should be employed in order to ensure that the trajectory tracking error remains as small as possible \citep{alcala2019lpv}.  

A wide variety of control strategies for autonomous vehicles (AVs) have already been proposed in literature \citep{amer2017modelling}. Model predictive control (MPC) is a widely used advanced control technique regarded as advantageous when it comes to control of AVs, as it allows for the direct treatment of actuator and state constraints as well as the incorporation of information regarding the future reference trajectory during the determination of the control signal, at the cost of high computational complexity \citep{yu2021model}.

The non-linearity of vehicle dynamics naturally leads to nonlinear optimization problems (NLP) when formulating a MPC strategy for AV control. The direct handling of such NLPs may be unfeasible due to its computational burden in conflict with real-time requirements \citep{Falcone2007}. Several strategies to overcome the limitations imposed by the NLPs in nonlinear model predictive control (NMPC) have already been studied and usually rely on replacing the original NLP by a sequence of simpler optimization problems (OPs)\citep{cannon2004efficient}, with differences mostly due to the model employed for predictions and to the consideration of coupled or uncoupled longitudinal and lateral dynamics for vehicle control \citep{amer2017modelling}. 

The authors of \citep{alcala2019lpv} employ a hierarchical control strategy for AV control, using a  Linear Parameter Varying (LPV) MPC formulation, with the reference trajectory as scheduling variables. They consider the kinematic bicycle model for predictions and use it to formulate the LPV-MPC that acts as a high-level controller. A LPV-LQR controller, designed using the dynamic bicycle model (DBM) for the vehicle, is used as a low-level controller. The low-level controller is responsible for ensuring that the linear acceleration and steering angle commands issued by the high level controller are tracked properly. In a similar fashion, in \citep{pang2022practical} a MPC strategy that relies on the linearization of the kinematic bicycle model around a reference trajectory is proposed and evaluated using a reduced scale AV. 

The RTI scheme has already been used in the development of several MPC methods for AVs control. The work presented in \citep{lopez2021predictive} describes the implementation of a NMPC strategy based on RTI that uses a simplified version of the DBM for predictions. The authors of \citep{vaskov2023friction} present a stochastic NMPC strategy for AVs based on RTI aiming at handling and adapting to the uncertainties in tire-road interactions, also using the DBM for predictions. A NMPC scheme for AVs based on the RTI scheme was proposed and experimentally evaluated, using a reduced scale vehicle, in \citep{quirynen2020integrated} demonstrating the effectiveness of the method in a real situation.

The MPC strategies based on the linearization of the system's dynamics along a reference trajectory are justified by its low computational burden, at the cost of lower prediction accuracy and control quality. On the other hand, as shown in \citep{gros2020linear}, RTI-based NMPC is expected to perform significantly better than methods based on linearization along a reference trajectory, at a very similar computation cost. 

In this context, the objective of this work is to describe the formulation of a NMPC strategy based on the Real-Time Iteration (RTI) scheme, originally proposed in \citep{diehl2001real}. The present formulation differs from the original one as it employs an implicit discretization of the system's dynamics with the objective of achieving greater prediction accuracy at a low computation cost for stiff dynamical systems, such as autonomous vehicles. The method proposed in this work differs from those found in literature by the fact that it derives the sequence of QPs solved in the RTI scheme using an implicit discretization of the system's dynamics. For stiff dynamical systems the employment of implicit discretization schemes prior to numerical integration allows for the usage of larger time steps and for greater accuracy when compared to the use of explicit discretization schemes. In practice greater time steps may render more computationally tractable NMPC controllers with a larger prediction time horizon.

The rest article is organized as follows: Section \ref{sec:methods} describes the formulation for the NMPC strategy considered in this study. Section \ref{sec:vehicle-dynamics} provides a background on vehicle dynamics and presents a detailed model used for simulation and a simplified model used for the prediction step in NPMC. Section \ref{sec:study-case} describes the case study simulation scenario considered for the evaluation of the NMPC technique and discusses the results. Finally, the conclusion of the paper is presented in section \ref{sec:conclusion}.

\section{Implicit Real-Time Iteration Nonlinear Model Predictive Control }\label{sec:methods}

This section presents the derivation of the sequence of OPs to be solved to realize a RTI-NMPC strategy using the implicit Euler method for the discretization of the differential equations describing the system's dynamics.

It is assumed that the system's dynamics can be described by a set of first-order nonlinear differential equations in the form given by equation \eqref{ode}.

\begin{equation}\label{ode}
    \dot{x} = f(x(t),u(t))
\end{equation}

The implicit Euler discretization method is characterized by a particular approximation of the differential equation \eqref{ode}. This approximation is presented in \eqref{ode-integral-approximation}. 

\begin{equation} \label{ode-integral-approximation}
   x(t) - x(t-\Delta t)  \approx f(x(t),u(t)) \Delta t
\end{equation}

For practical implementations of digital controllers based on NMPC, it is usually not possible to compute a control signal in a short period of time, hence it is necessary to modify equation \eqref{ode-integral-approximation} to account for the time-delay induced by the computation. In this work it is assumed that the there is a time-delay of one sampling period between the acquisition of a measurement and the update of the control signal. To account for this fact, equation \eqref{ode-integral-approximation} can be modified to produce equation \eqref{ode-integral-approximation-delay}.

\begin{equation} \label{ode-integral-approximation-delay}
   x(t) - x(t-\Delta t)  \approx f(x(t),u(t-\Delta t)) \Delta t
\end{equation}

To define the OP to be solved when realizing the NMPC strategy it is necessary to define the variables involved in such problem. The sequence of state predictions is described by the state prediction vector presented in equation \eqref{state-prediction-vector}. The notation $\hat{x}(t+i|t)$ stands for the predicted state at time $t+i$ using the information available at time $t$. The same notation is used for the reference trajectory and for the control signal increment, expressed in equations \eqref{reference-vector} and \eqref{control-increment-vector}, respectively.

\begin{equation}\label{state-prediction-vector}
    \boldsymbol{\hat{x}} = 
    \begin{bmatrix}
    \hat{x}(t|t) \\
    ... \\
    \hat{x}(t+N|t) \\
    \end{bmatrix}
\end{equation}

\begin{equation}\label{reference-vector}
    \boldsymbol{r} = 
    \begin{bmatrix}
    r(t|t) \\
    ... \\
    r(t+N|t) \\
    \end{bmatrix}
\end{equation}

\begin{equation}\label{control-increment-vector}
    \Delta\boldsymbol{u} = 
    \begin{bmatrix}
    u(t|t)-u(t-1|t) \\
    ... \\
    u(t+N|t)-u(t+N-1|t) \\
    \end{bmatrix}
\end{equation}

The discrete-time dynamics of the predictions obtained via the implicit Euler approximation can be described by equations \eqref{discrete-vector-field} and \eqref{discrete-vector-field-detail}.

\begin{equation}\label{discrete-vector-field}
    \begin{aligned}
        \hat{x}(t+i|t) = F(\hat{x}(t+i|t),\hat{x}(t+i-1|t),u(t+i-1|t))
    \end{aligned}
\end{equation}

\begin{equation}\label{discrete-vector-field-detail}
    \begin{aligned}
        \hat{x}(t+i|t) = \hat{x}(t+i-1|t) + f(\hat{x}(t+i|t),u(t+i-1|t)) \Delta t
    \end{aligned}
\end{equation}

A NLP associated with the NMPC formulation can then be defined as presented in equation \eqref{original-nlp}, in which $G(.)$ defines a set of nonlinear constraints on states and inputs, $x_m$ represents the latest measurement of the system's state and $u_l$ represents the latest control signal applied to the system. The matrix $C$ captures the relation between the internal states of the system and its output variables. Finally, $Q$ and $R$ are the weighting matrices of the cost function. A solution to this OP renders a sequence of state predictions associated with the optimal inputs along the prediction horizon.

As described in \citep{diehl2001real}, the RTI scheme consists of decomposing the NLP given by equation \eqref{original-nlp} into a sequence of QPs with linear constraints. At each sampling instant one QP from the sequence is then solved and the first sample of the resulting control signal sequence is applied to the system. Details regarding the convergence properties RTI can be found in \citep{diehl2001real}. 

\begin{equation}\label{original-nlp}
\begin{aligned}
\min_{\boldsymbol{\hat{x}},\Delta \boldsymbol{u}} \quad & [\boldsymbol{r}-C\boldsymbol{\hat{x}}]^T Q [\boldsymbol{r}-C\boldsymbol{\hat{x}}] + \Delta \boldsymbol{u}^T R \Delta \boldsymbol{u}\\
\textrm{s.t.} \quad & \hat{x}(t|t) = x_m \\
                    & u(t-1|t) = u_l \\
                    & u(t-N|t) = u(t-N-1|t) \\
                    & \hat{x}(t+i+1|t) = \\ \quad &F(\hat{x}(t+i|t),\hat{x}(t+i-1|t),u(t+i-1|t)) \\
                    & G(\hat{x}(t+i|t), u(t+i|t)) \leq 0
\end{aligned}
\end{equation}

The formulation of the sequence of quadratic problems starts by defining an initial guess for the solution of the OP. The guess for the state prediction sequence is denoted by $\boldsymbol{x}_g$ and the guess for the control inputs sequence is denoted by $\boldsymbol{u}_g$. A linear approximation for the sequence of predictions can then be derived using equation \eqref{implicit-linearized-dynamics}, in which $\delta\hat{x}(t+i|t) = \hat{x}(t+i|t) - x_g(t+i|t)$ and $\delta u(t+i|t) = u(t+i|t) - u_g(t+i|t)$ represents the deviations of the state prediction and of the predicted control input from their corresponding guesses. To simplify the notation, $F_g(.)$ is defined as in equation \eqref{F-implicit} and the matrices $A_0(.)$, $A_1(.)$ and $B(.)$ are defined by equations \eqref{A0-implicit}, \eqref{A1-implicit} and \eqref{B-implicit}, respectively.

\begin{equation}\label{F-implicit}
    F_g(t+i|t) =  F(x_g(t+i|t), x_g(t+i-1|t),u_g(t+i-1|t))
\end{equation}

\begin{equation}\label{implicit-linearized-dynamics}
    \begin{aligned}
            \hat{x}(t+i|t) &= F_g(t+i|t) + A_0(t+i|t)\delta\hat{x}(t+i-1|t) 
            \\ &+ A_1(t+i|t) \delta\hat{x}(t+i|t) 
            \\&+ B(t+i|t)\delta u(t+i-1|t)
    \end{aligned}
\end{equation}

\begin{equation}\label{A0-implicit}
    \begin{aligned}
        A_0(t+i|t) &= \quad \\ 
        \frac{\partial F}{\partial \hat{x}(t+i-1|t)}&(x_g(t+i|t),x_g(t+i-1|t), u_g(t+i-1|t))
    \end{aligned}
\end{equation}

\begin{equation}\label{A1-implicit}
    \begin{aligned}
        A_1(t+i|t) &= \quad \\
        \frac{\partial F}{\partial \hat{x}(t+i|t)}&(x_g(t+i|t), x_g(t+i-1|t),u_g(t+i-1|t))
    \end{aligned}
\end{equation}

\begin{equation}\label{B-implicit}
    \begin{aligned}
        B(t+i|t) &= \quad 
        \\ \frac{\partial F}{\partial u(t+i|t)}&(x_g(t+i|t), x_g(t+i-1|t),u_g(t+i-1|t))
    \end{aligned}
\end{equation}

By subtracting both sides of equation \eqref{implicit-linearized-dynamics} by $x_g(t+1|t)$, it is possible to define the discrete-time dynamics of the state predictions deviations from the guess as in equation \eqref{implicit-linearized-virtual-displacement-dynamics}. 

A similar procedure is adopted to provide a linear approximation for the nonlinear constraints defined by the function $G(.)$. The result of such procedure is presented in \eqref{linearized-inequality-contraint}, in which $G_g(t+i|t)$ is defined by equation \eqref{G-vector} and the matrices $D(.)$ and $E(.)$ are given by equations \eqref{D-matrix} and \eqref{E-matrix}, respectively.

\begin{equation}\label{implicit-linearized-virtual-displacement-dynamics}
    \begin{aligned}
            \delta\hat{x}(t+i|t) &= F_g(t+i|t) - x_g(t+i|t) \\
            &+ A_0(t+i|t)\delta\hat{x}(t+i-1|t) \\
            &+ A_1(t+i|t) \delta\hat{x}(t+i|t) \\ 
            &+ B(t+i|t)\delta u(t+i-1|t)
    \end{aligned}
\end{equation}

\begin{equation}\label{linearized-inequality-contraint}
    \begin{aligned}
            G(\hat{x}(t+i|t), u(t+i|t)) &\approx G_g(t+i|t) + D(t+i|t)\delta\hat{x}(t+i|t) \\
            &+ E(t+i|t)\delta u(t+i-1|t)
    \end{aligned}
\end{equation}

\begin{equation}\label{G-vector}
    G_g(t+i|t) = G(x_g(t+i|t), u_g(t+i-1|t))
\end{equation}

\begin{equation}\label{D-matrix}
    D(t+i|t) = \frac{\partial G}{\partial \hat{x}}(x_g(t+i|t), u_g(t+i-1|t))
\end{equation}

\begin{equation}\label{E-matrix}
    E(t+i|t) = \frac{\partial G}{\partial u}(x_g(t+i|t), u_g(t+i-1|t))
\end{equation}

By replacing $\boldsymbol{\hat{x}}$ and $\boldsymbol{u}$ by $\boldsymbol{\hat{x}_g} + \delta\boldsymbol{\hat{x}}$ and $\boldsymbol{u_g} + \delta\boldsymbol{u}$, respectively, in the original NLP from equation \eqref{original-nlp}, as well as replacing the constraints by their respective approximations, it is possible to define a QP with linear constraints as in equation \eqref{implicit-optimization-problem}. Therefore the solution of the proposed OP is a sequence of state and inputs deviations, $\delta\boldsymbol{\hat{x}}$ and $\delta\boldsymbol{u}$, which can be combined with the initial solution guess to produce the actual states prediction and control inputs sequence.

\begin{equation}\label{implicit-optimization-problem}
\begin{aligned}
\min_{\delta\boldsymbol{\hat{x}},\delta\Delta \boldsymbol{u}} \quad & [\boldsymbol{r}-C(\boldsymbol{x_g} + \delta\boldsymbol{\hat{x}})]^T Q [\boldsymbol{r}-C(\boldsymbol{x_g} + \delta\boldsymbol{\hat{x}})] \\ 
& + (\Delta \boldsymbol{u_g} + \delta\Delta \boldsymbol{u_g})^T R (\Delta \boldsymbol{u_g} + \delta\Delta \boldsymbol{u_g})\\
\textrm{s.t.} \quad & x_g(t|t) + \delta\hat{x}(t|t) = x_m \\
                    & u_g(t|t) + \delta u_g(t|t) = u_l \\
                    & u_g(t-N|t) + \delta u(t-N|t) = u_g(t-N-1|t)  \\ 
                    &+ \delta u(t-N|t) \\
                    & \delta\hat{x}(t+i|t) = F_g(t+i|t) - x_g(t+i|t) \\
                    &+ A_0(t+i|t)\delta\hat{x}(t+i-1|t) \\ &+ A_1(t+i|t) \delta\hat{x}(t+i|t) + B(t+i|t)\delta u(t+i-1|t) \\
                    & G_g(t+i|t) + D(t+i|t)\delta\hat{x}(t+i|t) \\ 
                    &+ E(t+i|t)\delta u(t+i-1|t) \leq 0
\end{aligned}
\end{equation}

Equation \eqref{implicit-optimization-problem}, then, describes the final QP that is solved at each sampling instant to produce control signals. 

In the next few sections this method is going to be employed to realize a NMPC control strategy for AVs.

\section{Vehicle Dynamics Models}\label{sec:vehicle-dynamics}

A defining characteristic of control strategies based on MPC is the use of a model of the system to be controlled to obtain predictions regarding its future states. In this section two vehicle dynamics models are going to be presented. The first is a four-wheel model with nonlinear tire force characteristics, which will be used to represent the plant. The second is a simplified model based on the DBM with linear tire force characteristics, which will be used to obtain the predictions for the MPC. Both models are based on the material presented in \citep{rajamani2011vehicle}.

\subsection{Simulation Model}\label{sec:simulation-model}

The four-wheel vehicle dynamics model presented in \citep{rajamani2011vehicle} can be described by the differential equations provided in equation \eqref{vehicle-dynamics}, in which $p_x$ and $p_y$ represent the vehicle's position in the inertial reference frame, $\psi$ represents the orientation of the vehicle with respect to the inertial reference frame, $v_x$ and $v_y$ are the vehicle's longitudinal and lateral velocities with respect to the body reference frame, $\omega$ is the angular velocity of the body reference frame with respect to the inertial reference frame and $\delta_l$ and $\delta_r$ are the left and right wheel steering angles. Also, $I$ stands for the moment of inertia of the vehicle about the $z$ axis, $m$ stands for the vehicle's mass, $b_{lon}$ and $b_{lat}$ stands for the drag coefficient in the longitudinal and lateral directions, respectively. The quantity $v_w$ represents the wind absolute velocity and $\alpha_w$ is the orientation of the wind velocity with respect to the inertial reference frame.

\begin{equation}\label{vehicle-dynamics}
    \begin{aligned}
        \dot{p}_x &= v_x cos(\psi) - v_y sin(\psi) \\
        \dot{p}_y &= v_x sin(\psi) - v_y cos(\psi) \\ 
        \dot{\psi} &= \omega \\
        \dot{v}_x &= \omega v_y + \frac{1}{m} [ F_{x,fl} cos(\delta_l) + F_{xfr} cos(\delta_r) \\ 
        &- F_{y,fl}sin(\delta_l)- F_{y,fr}sin(\delta_r) + F_{x,rl} + F_{x,rr}\\ &- b_{lon}(v_x + v_w cos(\alpha_w - \psi))^2]\\
        \dot{v}_y &= -\omega v_x + \frac{1}{m} [ F_{x,fl} sin(\delta_l) + F_{xfr} sin(\delta_r) \\ 
        &+F_{y,fl}cos(\delta_l)+ F_{y,fr}cos(\delta_r) + F_{y,rl} + F_{y,rr}\\ &- b_{lat}(v_y + v_w sin(\alpha_w - \psi))^2]\\
        \dot{\omega} &= \frac{1}{I} [ (F_{x,fl} sin(\delta_l) + F_{xfr} sin(\delta_r) \\ 
        &+ F_{y,fl}cos(\delta_l)+F_{y,fr}cos(\delta_r))l_f - (F_{y,rl} + F_{y,rr})l_r\\ 
        & + (F_{x,fr} cos(\delta_r) - F_{xfl} cos(\delta_l) + F_{y,fl}sin(\delta_l) \\&
        - F_{y,fr}sin(\delta_r) - F_{x,rl} + F_{x,rr})t_w
    \end{aligned}
\end{equation}

The dynamics of the vehicle is mostly defined by the tire forces, represented as $F_{y,.}$ for lateral forces and $F_{x,.}$ for longitudinal forces. The $fl$, $fr$, $rl$ and $rr$ subscripts stand for front-left, front-right, rear-left and rear-right tires, respectively. The tire forces are computed based on the slip angle, defined in equation \eqref{slip-angle}, and the slip ratio, defined in equation \eqref{slip-ratio}. The model relating those quantities to their respective tire forces is the modified Dugoff model presented in \citep{bian2014dynamic} and omitted in this work for brevity.

\begin{equation}\label{slip-angle}
    \begin{aligned}
        \alpha_{f} &= \delta - tan^{-1}(\frac{v_y + l_f \omega}{v_x}) \\ 
        \alpha_{r} &= - tan^{-1}(\frac{v_y - l_r \omega}{v_x}) \\ 
    \end{aligned}
\end{equation}

\begin{equation}\label{slip-ratio}
    \sigma = \frac{r w - v_x}{v_x}
\end{equation}

It is necessary, also, to account for the wheel dynamics of the vehicle, which are given by equation \eqref{wheels-dynamics}, in which $J$ is the wheel inertia, $r$ is the wheel's effective radius and $u_{\tau}$ is the torque applied to the wheels, one of the control inputs for the system.

\begin{equation}\label{wheels-dynamics}
    \begin{aligned}
        \dot{w}_{fr} &= \frac{1}{J}(-r F_{x,fr}) \\
        \dot{w}_{fl} &= \frac{1}{J}(-r F_{x,fl}) \\
        \dot{w}_{rr} &= \frac{1}{J}(u_{\tau} -r F_{x,rr}) \\
        \dot{w}_{rl} &= \frac{1}{J}(u_{\tau} -r F_{x,rl}) \\
    \end{aligned}
\end{equation}

The steering angles of the right and left front wheel are given by the Ackerman geometry and a control input to the system is the rate of change of the reference average steering angle $\delta_c$, whose dynamics, along with the expressions for the Ackerman geometry, are given by equation \eqref{steering-dynamics}.

\begin{equation}\label{steering-dynamics}
    \begin{aligned}
        \dot{\delta}_c &= u_\delta \\
        R &= \frac{(l_f+l_r)}{tan(\delta_c)} \\
        \delta_r &= tan^{-1}(\frac{L}{R+t_w})\\
        \delta_l &= tan^{-1}(\frac{L}{R-t_w})
    \end{aligned}
\end{equation}

The use of this detailed model directly in an MPC formulation would render a great computational cost due to its high dimensionality and complexity. On the other hand, as analysed in \citep{zhou2023impact} a simplified model based on the DBM can provide an accurate approximation of the vehicle dynamics model while providing a smaller computational burden when solving the OPs for the MPC.

\subsection{Prediction Model}\label{sec:prediction-model}

In the DBM, as described in \citep{rajamani2011vehicle}, the dynamics of a four-wheel vehicle is approximated through an equivalent two-wheel vehicle, in which the steering angle of the front wheel is given by the average between the steering angles of both front wheels from the four-wheel vehicle. This dynamic model can be expressed by equation \eqref{bicycle-dynamics}, the wheels dynamics can be expressed by equation \eqref{bicycle-wheels-dynamics} and the definitions of the slip angle and the slip ratio remain the same as in the four wheel model.

Another approximation used to simplify the model is that of linear tire characteristics, which is usually sufficiently accurate for small slip angles and slip ratios. The longitudinal and lateral tire forces can then be computed based on equations \eqref{longitudinal-force} and \eqref{lateral-force}, respectively, in which $C_s$ stands for the tire longitudinal stiffness coefficient and $C_a$ is the tire lateral stiffness coefficient.

When considering actual tire characteristics, the forces saturate as the slip angle and slip ratios increase. To capture this phenomena in the proposed MPC, a constraint on the longitudinal and lateral forces, given by the friction circle, is imposed on the predictions, as presented in \eqref{friction-circle}, in which $\mu$ is the road friction coefficient and $g$ is the gravity acceleration.

\begin{equation}\label{bicycle-dynamics}
    \begin{aligned}
        \dot{p}_x &= v_x cos(\psi) - v_y sin(\psi) \\
        \dot{p}_y &= v_x sin(\psi) - v_y cos(\psi) \\ 
        \dot{\psi} &= \omega \\
        \dot{v}_x &= \omega v_y + \frac{1}{m} [ 2 F_{x,f} cos(\delta_c) - 2 F_{y,f}sin(\delta_c) + 2 F_{x,r} 
        \\ &- b_{lon}(v_x + v_w cos(\alpha_w - \psi))^2]\\
        \dot{v}_y &= -\omega v_x + \frac{1}{m} [ 2 F_{x,f} sin(\delta_c) 
        + 2F_{y,f}cos(\delta_c) + 2 F_{y,r
        }\\ &- b_{lat}(v_y + v_w sin(\alpha_w - \psi))^2]\\
        \dot{\omega} &= \frac{1}{I} [ (2F_{x,f} sin(\delta_c) 
        + 2 F_{y,f}cos(\delta_c))l_f - 2F_{y,r}l_r\\ 
    \end{aligned}
\end{equation}

\begin{equation}\label{bicycle-wheels-dynamics}
    \begin{aligned}
        \dot{w}_{f} &= \frac{1}{J}(-r F_{x,f}) \\
        \dot{w}_{r} &= \frac{1}{J}(u_{\tau} -r F_{x,r}) \\
    \end{aligned}
\end{equation}

\begin{equation}\label{longitudinal-force}
    F_{lon} = C_s \sigma
\end{equation}

\begin{equation}\label{lateral-force}
    F_{lat} = C_a \alpha
\end{equation}

\begin{equation}\label{friction-circle}
    F_{lat}^2 + F_{lon}^2 \leq (\mu mg)^2
\end{equation}

Due to its sufficient accuracy when it comes to modeling the actual vehicle dynamics \citep{zhou2023impact}, in this work this simplified DBM is considered for the computations of predictions in the proposed NMPC. 

\section{Case Study}\label{sec:study-case}

To evaluate the proposed approach for AV control using the RTI-NMPC the closed-loop system is evaluated in a simulation scenario. For comparison, a closed-loop system using an NMPC controller based on the direct solution of the NLP from equation \eqref{original-nlp} is also evaluated on the same simulation scenario. The models used for simulation and for the MPC predictions are those described in sections \ref{sec:simulation-model} and \ref{sec:prediction-model}, respectively. 

The results were acquired using a computer with an apple M2 max processor and 32GB of RAM and the solver used to realize the MPC was Ipopt.

\begin{table}[h]
\begin{center}
\caption{System Parameters.}\label{tb:parameters}
\begin{tabular}{cccc}
Parameter & Value \\\hline
$m$ & $200$ [$kg$]  \\
$I$ & $150$ [$kg.m^2$] \\ 
$l_r$ & $1.0$ [$m$] \\ 
$t_w$ & $0.6$ [$m$] \\ 
$l_f$ & $0.8$ [$m$] \\ 
$r$ & $0.6$ [$m$] \\ 
$J$ & $0.2$ [$kg.m^2$] \\
$b_{lon}$ & $0.01$ [$N s^2/m^2$] \\
$b_{lat}$ & $0.05$ [$N s^2/m^2$] \\
$C_s$ & $600000 [N] $ \\
$C_a$ & $250000$ [$N/rad$] \\
$\mu$ & $0.9$  \\
\end{tabular}
\end{center}
\end{table}

\subsection{Scenario Description}

The simulated scenario consists in the vehicle following a reference trajectory while subject to measurement noise and to wind disturbances modeled as integrated Gaussian noise, as such disturbances are related, in general, to adverse conditions for driving \citep{hou2019framework}. The vehicle parameters used for simulation are presented in table \ref{tb:parameters}.

The wind disturbance is modeled as integrated Gaussian noise with a mean of $2 m/s$ and a standard deviation of $1.5 m/s$. The resulting wind disturbance waveform in presented in figure \ref{fig:wind-disturbance}. The measurement noise is modeled as zero-mean Gaussian noise and the standard deviation associated with each state variable is presented in table \ref{tb:noise-parameters}. To reduce the impacts of measurement noise in the controller performance, a first-order Butterworth with cutoff frequency at $3.5 Hz$ was designed, discretized using the Tustin method and applied to all measurement signals.

The system's dynamics are discretized using a $0.04 s$ time step, equal to the sampling period. The prediction horizon considered is $15$ samples. The cost function is normalized by the minimum and maximum values expected for control action and reference trajectory values. Besides the constraint given by equation \eqref{friction-circle}, additional constraints are imposed to the maximum and minimum values of wheel torque, steering angle rate and steering angle. The minimum and maximum wheel torque, steering angle rate and steering values are $-300 Nm$ and $300 Nm$,  $-1.5 rad/s$ and $1.5 rad/s$, $-45^{\circ}$ and $45^{\circ}$ respectively. A relative weight, in the cost function, of $5$ is given to both the wheel torque control signal increment and steering angle rate control signal increment.

\begin{table}[h]
\begin{center}
\caption{Measurement Noise Parameters.}\label{tb:noise-parameters}
\begin{tabular}{cccc}
State Variable & Noise Distribution Standard Deviation \\\hline
$p_x$ & $0.05$ [$m$]  \\
$p_y$ & $0.05$ [$m$]  \\
$\psi$ & $0.01$ [$rad$]  \\
$v_x$ & $0.05$ [$m/s$]  \\
$v_y$ & $0.05$ [$m/s$]  \\
$\omega$ & $0.01$ [$rad/s$] \\ 
$\omega_{fr}$ & $0.01$ [$rad/s$] \\ 
$\omega_{fl}$ & $0.01$ [$rad/s$] \\ 
$\omega_{rr}$ & $0.01$ [$rad/s$] \\ 
$\omega_{rl}$ & $0.01$ [$rad/s$] \\ 
$\delta_c$ & $0.01$ [$rad$] \\
\end{tabular}
\end{center}
\end{table}

\subsection{Results}

This section presents the results of the RTI-NMPC controller evaluation in the proposed scenario. 

\begin{figure}
    \centering
    \begin{subfigure}[b]{0.4\textwidth}
         \centering
         \includegraphics[width=\textwidth]{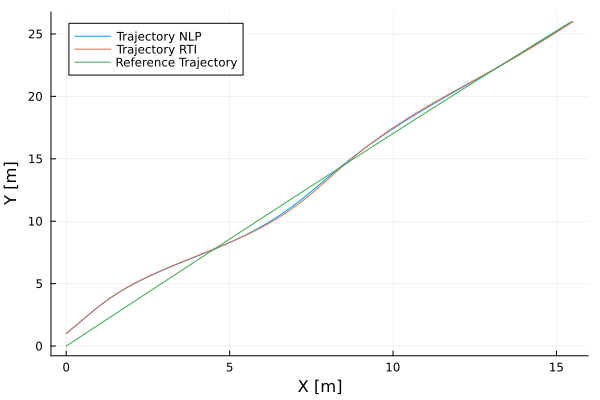}
         \caption{Simulated trajectories during the first $2s$ of the simulation}
         \label{fig:zoom-trajectory}
    \end{subfigure}
    \begin{subfigure}[b]{0.4\textwidth}
         \centering
         \includegraphics[width=\textwidth]{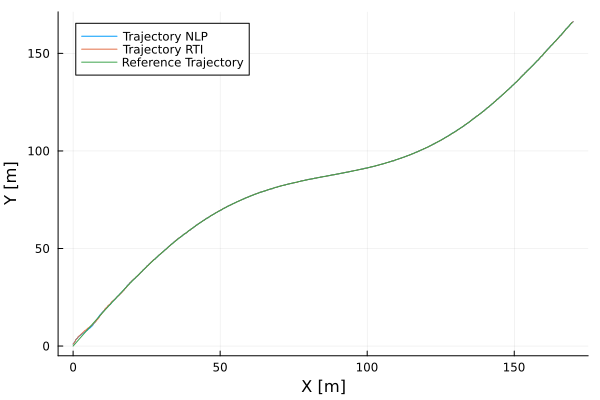}
         \caption{Simulated trajectories during $24s$ of the simulation}
         \label{fig:trajectory}
    \end{subfigure}
    \caption{Reference and simulated trajectories for the NMPC based on RTI and direct NLP solution.}
\end{figure}

Figures \ref{fig:zoom-trajectory} and \ref{fig:trajectory} show the reference and simulated trajectories during the closed loop simulations using the RTI and direct NLP based controllers, demonstrating that both controllers are able to successfuly track the proposed trajectory.

\begin{figure}
     \centering
     \begin{subfigure}[b]{0.42\textwidth}
         \centering
         \includegraphics[width=\textwidth]{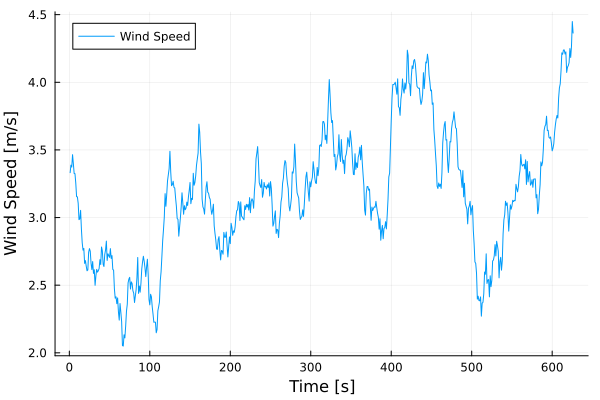}
         \caption{Wind disturbance}
         \label{fig:wind-disturbance}
     \end{subfigure}
     \hfill
     \begin{subfigure}[b]{0.42\textwidth}
         \centering
         \includegraphics[width=\textwidth]{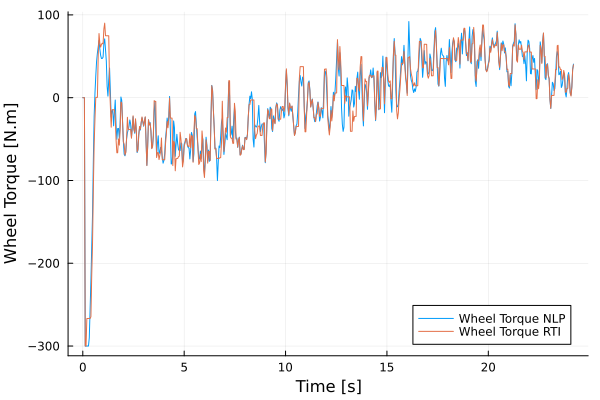}
         \caption{Wheel torque control signals}
         \label{fig:wheel-torque}
     \end{subfigure}
     \hfill
          \begin{subfigure}[b]{0.42\textwidth}
         \centering
         \includegraphics[width=\textwidth]{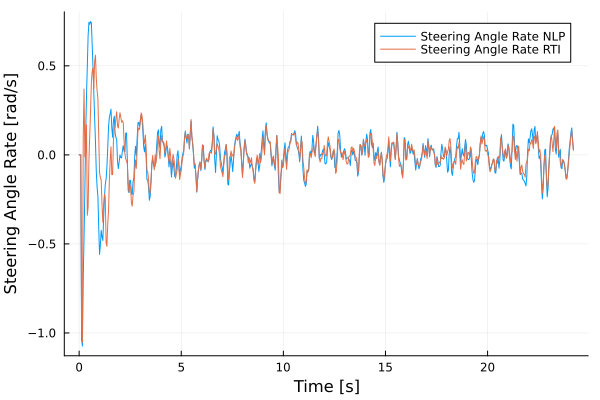}
         \caption{Steering angle rate control signals}
         \label{fig:steering-angle-rate}
     \end{subfigure}
     \hfill
          \hfill
          \begin{subfigure}[b]{0.42\textwidth}
         \centering
         \includegraphics[width=\textwidth]{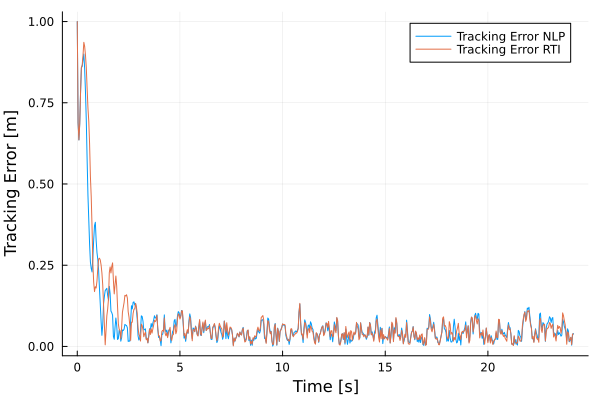}
         \caption{Tracking error}
         \label{fig:tracking-error}
     \end{subfigure}
     \hfill
    \caption{Simulation results for the RTI and direct NLP controllers: (a) Wind disturbance for the proposed evaluation scenario. (b) Wheel torque resulting. (c) Steering angle rate. (d) Tracking error.}
\end{figure}

Figures \ref{fig:wheel-torque} and \ref{fig:steering-angle-rate} present the wheel torque and steering angle control signals from both controllers. In both figures it is possible to visualize the high-frequency oscillations as a response to measurement noise, showing that the two controllers are equally sensitive to noise and produce very similar control signals. The tracking error for each controller is presented in figure \ref{fig:tracking-error}, which reinforces the effectiveness of both controllers when it comes to tracking a desired trajectory and also indicates that the controller based on the direct solution of the NLP does not lead, necessarily, to superior performance. To properly capture the transient behavior of the closed-loop system, the initial tracking error, given by the difference between the initial state of the system and the first point of the reference trajectory, was set to $1m$.

The fact that the control signals, shown in figures \ref{fig:wheel-torque} and \ref{fig:steering-angle-rate}, as well as the tracking error, shown in figure \ref{fig:tracking-error}, for both controllers do not present significant differences presents itself as an evidence that the solutions for the NMPC problem obtained using the RTI scheme successfully approximate the solutions obtained via nonlinear optimization methods. 

\begin{figure}
     \centering
     \begin{subfigure}[b]{0.4\textwidth}
         \centering
         \includegraphics[width=\textwidth]{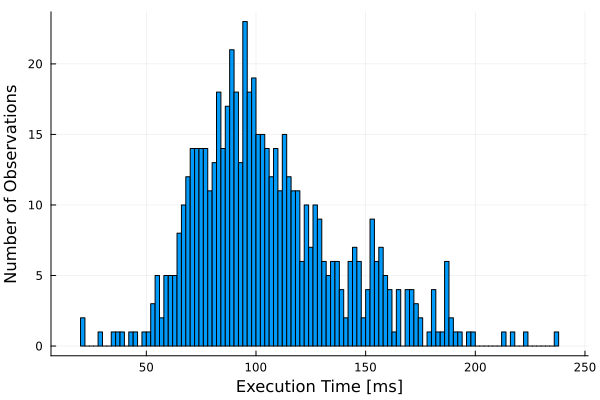}
         \caption{direct NLP solution}
         \label{fig:execution-time-nlp}
     \end{subfigure}
     \hfill
     \begin{subfigure}[b]{0.4\textwidth}
         \centering
         \includegraphics[width=\textwidth]{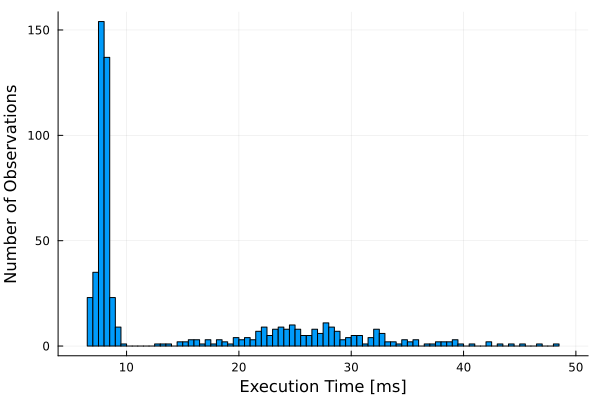}
         \caption{RTI}
         \label{fig:execution-time-rti}
     \end{subfigure}
     \hfill
    \caption{Execution time observations for the NMPC based on the direct NLP solution and on the RTI scheme}
\end{figure}

The results presented in figures \ref{fig:execution-time-nlp} and \ref{fig:execution-time-rti} stand for the execution times observed for the controller based on the direct NLP solution and for that based on the RTI scheme, respectively. It is clear that the practical realization of the controller based on the direct NLP solution in unfeasible, as all execution time observations for this controller are greater than the controller's sampling period, at an average of approximately $106 ms$,. On the other hand, the execution times observed for the controller that uses the RTI scheme are mostly below $40 ms$, the controller's sampling period, and have an average of approximately $15 ms$, rendering the controller realizable. Although the variability in execution times is inevitable due to the iterative nature of optimization algorithms, in a practical scenario, for the few cases in which solver cannot find a solution in time, it is still possible to use sub-optimal solution to obtain the control signal.

\section{Conclusion}\label{sec:conclusion}

In this work a strategy for trajectory tracking control of AVs based on the RTI scheme for NMPC considering an implicit discretization method was presented and a derivation of the quadratic sub-problems that composes the RTI scheme using the implicit Euler discretization method, specially useful when dealing with stiff dynamical systems, was provided. The ability of the proposed scheme to track a given trajectory was verified in simulation considering a detailed four-wheel vehicle dynamics model considering the effects of wind disturbances and measurement noise.

The effectiveness of the approach was evaluated based on its ability to track the reference trajectory as well as on its computational performance. The proposed strategy was also compared with another NMPC controller based on the direct solution of a nonlinear optimization problem. The presented results indicate that the proposed RTI-MPC strategy for trajectory tracking control of AVs is indeed effective. The trajectory tracking performance was satisfactory despite the presence of disturbances and modeling errors. The execution time observed when using the proposed were, on average, 7 times smaller than those observed when using a controller that solves the NMPC problem via direct nonlinear optimization, evidencing the computational efficiency associated with this strategy for nonlinear model predictive control.

\begin{ack}
The present work was realized with the support from CAPES, a Brazilian Government entity for human resources development. 
Julio E. Normey-Rico thanks the CNPq project 304032/2019-0.
\end{ack}

\bibliography{ifacconf}             
                                                   







\end{document}